\documentclass[sigconf]{acmart}
\AtBeginDocument{%
  \providecommand\BibTeX{{%
    \normalfont B\kern-0.5em{\scshape i\kern-0.25em b}\kern-0.8em\TeX}}}

\setcopyright{acmcopyright}
\copyrightyear{2021}
\acmYear{2021}
\acmDOI{10.1145/1122445.1122456}
\renewcommand{\baselinestretch}{0.95}
\acmConference[]{Proceedings of KDD DSHealth}{August 14-18, 2021}{}
\acmBooktitle{Proceedings of KDD DSHealth '21}
\acmPrice{15.00}
\acmISBN{978-1-4503-XXXX-X/18/06}
\usepackage{float}
\usepackage{titlesec}
\titlespacing{\section}{0pt}{5pt plus 0pt minus 2pt}{1pt plus 2pt minus 1pt}
\titlespacing{\subsection}{0pt}{5pt plus 0pt minus 2pt}{1pt plus 2pt minus 1pt}

\begin{document}

\title{GAN-based Data Augmentation for Chest X-ray Classification}

\author{Shobhita Sundaram}
\authornote{Both authors contributed equally to this research.}
\email{shobhita@mit.edu}
\author{Neha Hulkund}

\authornotemark[1]
\email{nhulkund@mit.edu}
\affiliation{%
  \institution{Massachusetts Institute of Technology}
  \streetaddress{77 Massachusetts Ave}
  \city{Cambridge}
  \state{Massachusetts}
  \country{USA}
}
\renewcommand{\shortauthors}{Sundaram and Hulkund, et al.}

\begin{abstract}
A common problem in computer vision -- particularly in medical applications -- is a lack of sufficiently diverse, large sets of training data. These datasets often suffer from severe class imbalance. As a result, networks often overfit and are unable to generalize to novel examples. Generative Adversarial Networks (GANs) offer a novel method of synthetic data augmentation. In this work, we evaluate the use of GAN- based data augmentation to artificially expand the CheXpert dataset of chest radiographs. We compare performance to traditional augmentation and find that GAN-based augmentation leads to higher downstream performance for underrepresented classes. Furthermore, we see that this result is pronounced in low data regimens. This suggests that GAN-based augmentation a promising area of research to improve network performance when data collection is prohibitively expensive.

\end{abstract}



\maketitle

\section{Introduction}
Convolutional Neural Networks (CNNs) have demonstrated tremendous success in recent years for visual tasks such as object recognition and segmentation. The application of such networks to clinical tasks -- including the segmentation of organs/pathologies \cite{segmentationSurvey}, and classification of medical images \cite{imagingSurvey} \cite{imagingSurvey2}-- promises to augment medical decision-making.

However, large amounts of labeled data are required to train high-performing CNNs. A lack of sufficiently large, diverse sets of training data typically results in models that overfit on the training data \cite{overfitting} and generalize poorly \cite{stutz2019disentangling}. Labelling medical imaging data is both expensive and time-consuming; lack of sufficiently diverse labelled training data is one of the major barriers to developing models fit for clinical use. Furthermore, medical data often suffers from the problem of class imbalance \cite{japkowicz2002class}, where samples of one pathology may be far more prevalent than others, leading to biased models \cite{krawczyk2016learning}. 

In this work, we investigate whether incorporating data generated by Generative Adversarial Networks (GANs) into training data as a data augmentation technique can improve the efficacy of Deep Neural Networks in diagnosing lung diseases from chest radiographs. While GAN augmentation requires an additional trained network -- as opposed to standard augmentation techniques -- our work demonstrates that this technique leads to performance improvements, which are particularly important in high stakes clinical decision-making.

While several prior works have tested the efficacy of GAN-based data augmentation, there are limited studies that compare effectiveness of GAN augmentation with more traditional methods of fixing class imbalance \cite{Sampath2020ASO}. We aim to bridge this gap by exploring the problem of class imbalance across different data regimens, comparing the performance of traditional and GAN data augmentation methods.

The contributions of this paper are:
\begin{itemize}
    \itemsep-.2em 
    \item We demonstrate the efficacy of GAN-based data augmentation compared to standard data augmentation and no augmentation techniques in correcting class imbalances that are found in the Stanford CheXpert dataset.
    \item We show that GAN data augmentation is most effective when used with small, significantly imbalanced datasets, and has limited impact for large datasets.
\end{itemize}

\section{Related Work} 
GAN data augmentation has been found useful for diversifying datasets by producing novel samples \cite{GoodfellowGAN} \cite{Shorten2019ASO}. Sanfort et al. and Rashid et al. found that using CycleGAN for augmentation in segmentation and classification tasks improved performance significantly, demonstrating the potential of GAN-based data augmentation for medical applications \cite{CycleGAN_CT_scan}\cite{Rashid_skin_lesion_GAN_augmentation}.

GAN data augmentation has been used to correct class imbalance with moderate success on imbalanced MNIST and CIFAR datasets using balancing GANS (BAGANs) \cite{Mariani2018BAGANDA}, as well as brain tumor datasets \cite{Qasim2020RedGANAC}. Further works have found that synthetic data augmentation for class imbalance is more effective for low data regimens, where the model suffers most from overfitting and lack of generalization \cite{antoniou2018data}. As far as we know this finding has not been studied in medical datasets, nor systematically compared to standard augmentation techniques. 

\section{CheXpert Dataset} 
\begin{figure}
\centering
\includegraphics[width=1\linewidth]{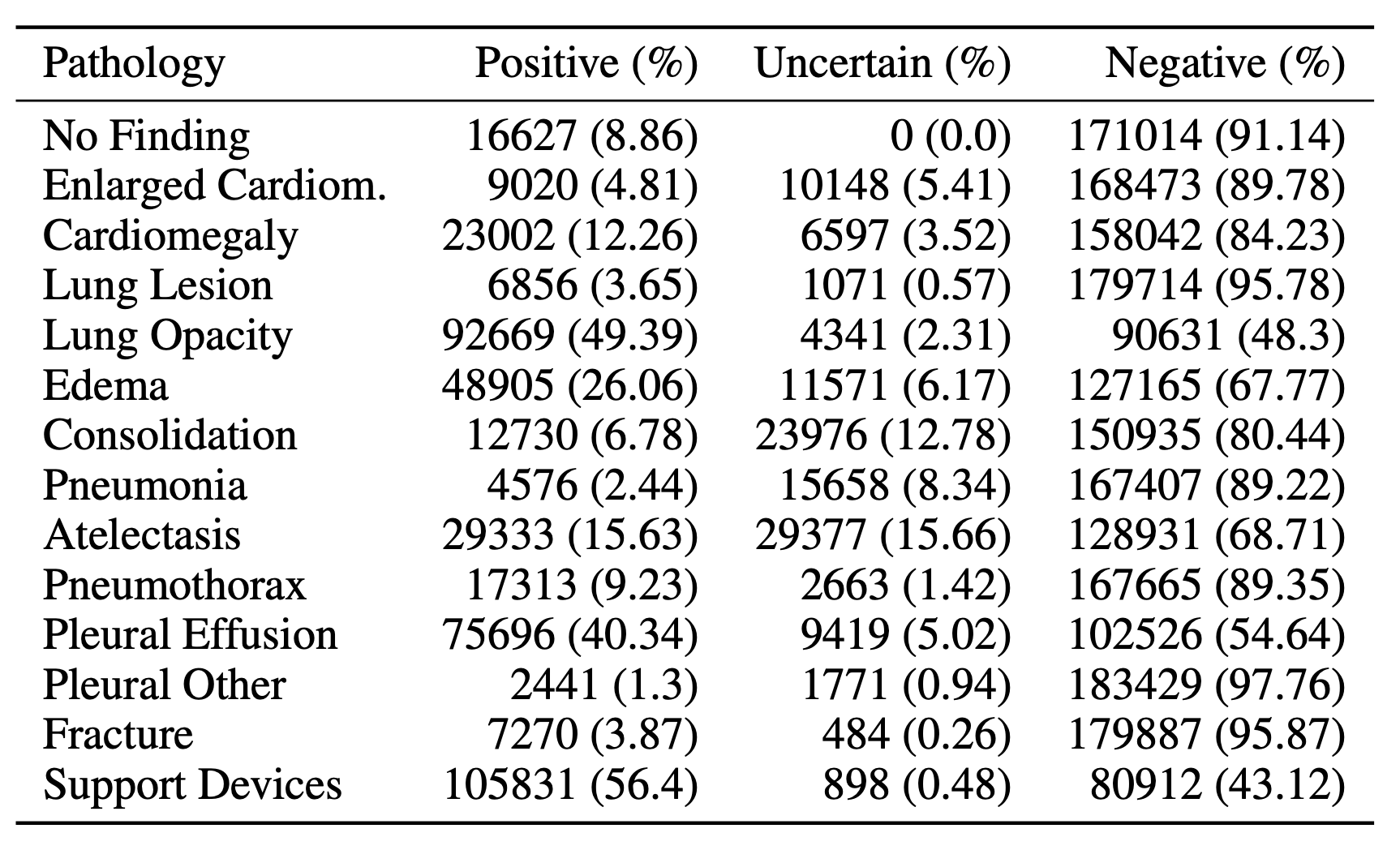}
\caption{Pathologies Present in Dataset \cite{Irvin2019CheXpertAL}}
\label{pathologies}
\end{figure}

Our experiment uses CheXpert, a publicly available dataset from Stanford Hospital of 224,316 frontal and lateral chest radiographs of 65,240 patients \cite{Irvin2019CheXpertAL}. Each radiograph is labeled with a $14$-element vector, where each element indicates the label for a particular pathology in the image; $1$ corresponds to positive, $0$ to negative, and $-1$ to uncertain. The different pathologies and their proportions in the dataset are detailed in Fig. \ref{pathologies}. 

We employ a label-smoothing technique of mapping the uncertainty labels to $1$, previously shown by Pham et al to improve performance in the dataset  \cite{Pham2021InterpretingCX}. Further preprocessing includes resizing the images to $224$ by $224$ pixels, performing a center crop for uneven aspect ratios, and scaling pixel values to [$-1024$, $1024$] \cite{cohen2020limits}. 

As shown in Fig. \ref{pathologies}, the CheXpert data suffers from significant class imbalance. We perform data augmentation on three pathologies which each make up less than 5\% of the dataset: Lung Lesion, Pleural Other, and Fracture. 

To examine the augmentation efficacy over different sizes of datasets, we perform all experiments over the following subsets of the dataset: 1\%, 10\%, 50\%, and 100\%, further detailed in Section 5.1.

\begin{table}
\small
\begin{center}
 \begin{tabular}{p{1.75cm}p{1cm}p{1.4cm}p{1.3cm}p{1.3cm}}
 \hline
 Dataset Size & Lung Lesion  & Pleural Other  & Fracture     & Augmented Dataset 
 \\ [0.5ex] 
 \hline
 2050 (1\%) & 500 & 550 & 500 & 3600 \\ 
 20371 (10\%) & 3500 & 4000 & 3500 & 49829\\
 100850 (50\%) & 15000 & 17000 & 15000 & 147850\\
 201391 (100\%) & 30000 & 35000 & 30000 & 296391 \\ [1ex]
 \hline
\end{tabular}
\end{center}
\caption{Number of generated images for each underrepresented class across different data regimens for GAN-augmentation}
\label{dataset_size}
\end{table}


\section{Approach}
\begin{figure*}[h]
\centering
\includegraphics[width=0.85\linewidth]{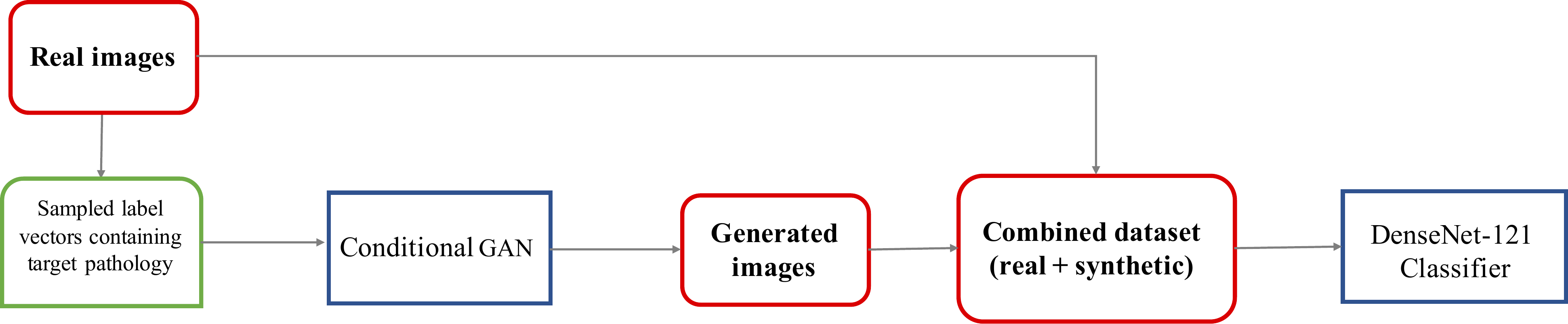}
\caption{Illustration of the proposed GAN-based data augmentation method. Our method combines GAN-generated images with real images of underrepresented pathologies to produce an augmented dataset that can be used for training.}
\label{fig:flowchart}
\end{figure*}

Our approach consists of generating GAN images in which at least one of the $3$ underrepresented pathologies are present. These GAN-generated images are then integrated with the original CheXpert data subset, such that the class imbalance is reduced. The process is shown in Fig \ref{fig:flowchart}. We evaluate the relative efficacy of training on GAN-augmented data by comparing to separate models trained with standard data augmentation and no data augmentation.

\subsection{Classification model} 
In previous works, the DenseNet-121 architecture has achieved a high performance on chest X-ray classification, as demonstrated in Rajpurkar et al and Pham et al \cite{Rajpurkar2018DeepLF} \cite{Densenet_for_XRay}. We transfer train TorchXRayVision's DenseNet-121, pretrained on ImageNet, for all experiments  \cite{cohen2020limits}. For each data augmentation procedure (refer to $4.2$), we train using $\{1\%, 10\%, 50\%, 100\%\}$ subsets of the full CheXpert dataset. 

\begin{figure}[H]
\centering
\includegraphics[width=.85\linewidth]{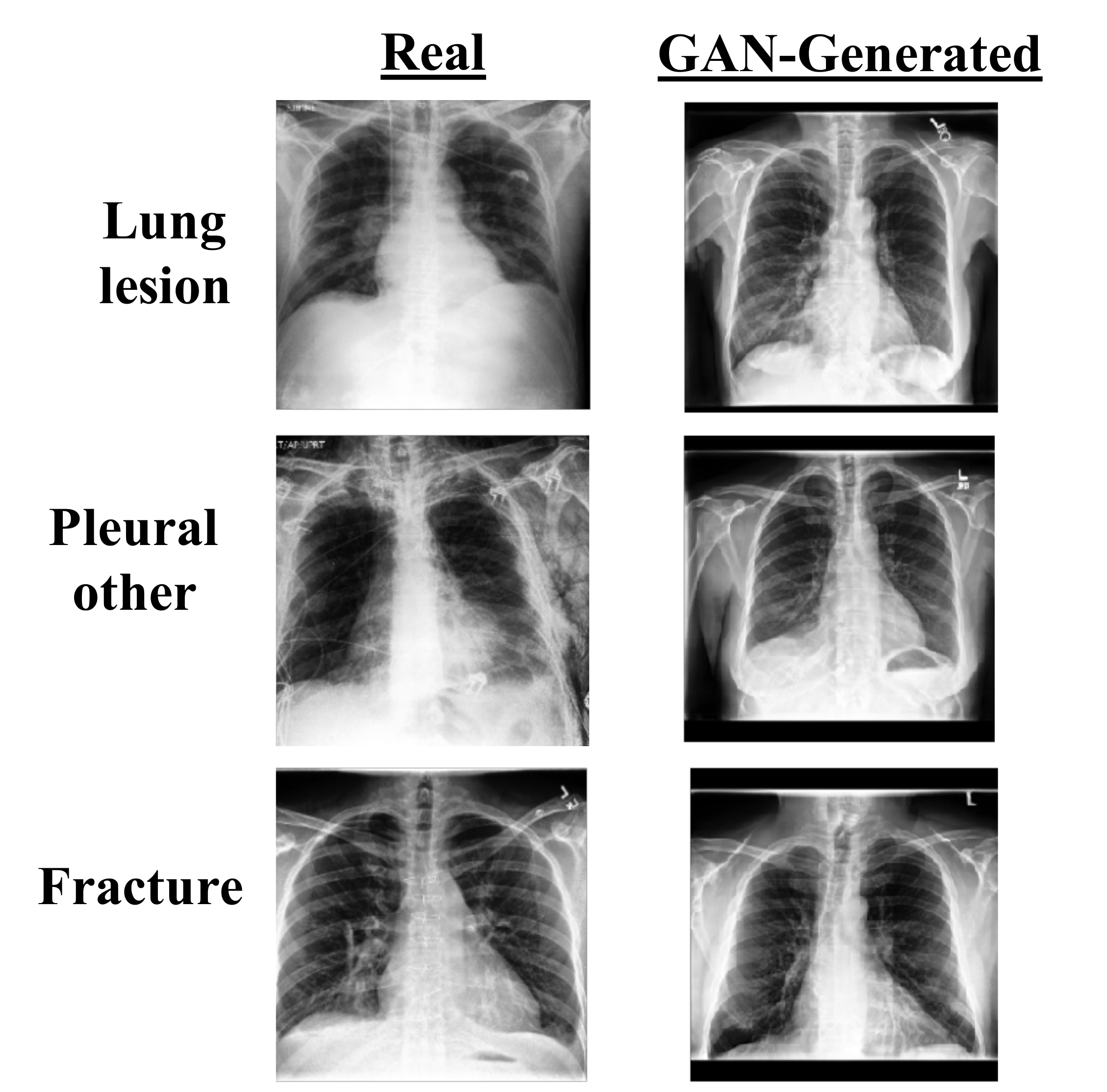}
\caption{Sample synthetic chest X-rays generated by a pretrained GAN for underrepresented classes in CheXpert}
\label{fig:GAN_imgs}
\end{figure}

\begin{table*}
\begin{center}
\begin{tabular}{{p{2cm}p{3cm}p{3cm}p{3.5cm}p{3cm}}}
 \hline
 Dataset Size & Pathology & No Augmentation & Standard Augmentation & GAN Augmentation
 \\ [0.5ex] 
 \hline
  & Lung Lesion & 0.727 & 0.728 & 0.758 \\
 1\% & Pleural Other & 0.566 & 0.550 & 0.594 \\
  & Fracture & 0.583 & 0.601 & 0.656 \\ 
 \hline
  & Lung Lesion & 0.790 & 0.796 & 0.803 \\
  10\% & Pleural Other & 0.632 & 0.655 & 0.670 \\
   & Fracture & 0.700 & 0.723 & 0.742 \\
  \hline
   & Lung Lesion & 0.826 & 0.822 & 0.828 \\
   50\% & Pleural Other & 0.710 & 0.696 & 0.706 \\
    & Fracture & 0.789 & 0.780 & 0.793 \\
  \hline
   & Lung Lesion & 0.835 & 0.832 & 0.834 \\
   100\% & Pleural Other & 0.721 & 0.712 & 0.727 \\
    & Fracture & 0.811 & 0.793 & 0.807 \\
  [1ex]
 
 \hline
\end{tabular}
\end{center}
\caption{AUC performance across augmentation techniques and dataset regimens for augmented classes \label{AUC_table}}
\end{table*}

\subsection{Data augmentation frameworks} 
\paragraph{\noindent{\textbf{GAN-based augmentation}}}Our proposed data augmentation framework employs GANs to generate synthetic images of underrepresented pathologies, thus correcting for class imbalance in the training dataset.

We use a Conditional GAN with mirrored structures for the generator and discriminator, that was progressively pretrained on chest X-rays from the CheXpert dataset as described in \cite{conditionalGAN}. The trained GAN generates radiographs that are difficult to distinguish from real images \cite{conditionalGAN}. The Conditional GAN takes as input a length-$14$ label vector $\{x_1, x_2,...,x_{14}\}$, where $x_i = 1$ if pathology $i$ is present, and $x_i = 0$ otherwise (note that if $x_0=1$ then the classification is "No Finding"). As output, it generates a chest X-ray with the corresponding pathologies present. Note that this gives us precise control over the exact label vector for each generated image.

Given an underrepresented pathology $p$, $L_p$ is the subset of label vectors in the real dataset in which $p$ is present (i.e. all label vectors for which $x_p=1$). We uniformly sample with replacement from $L_p$. Each sampled label vector is inputted to the GAN to generate a synthetic chest X-ray in which $p$ is present. Note that a synthetic image with $p$ present may contain other pathologies as well, to preserve any co-occuring relationships between pathologies that may be important towards classification. We then combine the set of generated images with the original image dataset to produce the more balanced augmented dataset. Sample synthetic images are presented alongside real X-rays in Fig. \ref{fig:GAN_imgs}

For each of the underrepresented pathologies (lung lesion, pleural other, fracture) we generate the minimum number of synthetic images such that each pathology is represented in at least $15\%$ of the combined augmented dataset. We choose the $15\%$ threshold because the majority of CheXpert pathologies are present in $5-26\%$ of images; thus $15\%$ representation enables a more balanced dataset.

The exact number of images generated for each training data subset is shown in Table \ref{dataset_size}.

\paragraph{\textbf{Standard data augmentation}} 
As a benchmark comparison to no augmentation and GAN augmentation, we apply  standard data augmentation techniques of randomly vertical/horizontal flipping a subset of the X-rays. We use these transformations because they have been shown to lead to performance gains for the chest X-ray datasets in previous works \cite{Stephen2019AnED}. In each epoch of the training process, we apply these transforms on-the-fly with probability $0.5$ to each image in the batch. 
\paragraph{\textbf{No data augmentation}} 
We additionally train DenseNet-121 without any augmentations. Training and evaluating this network creates a baseline with which to compare the networks trained on GAN-augmented data and standard-augmented data.
\section{Results} 
\subsection{Experimental setup}$ $ We train DenseNet-121 for a $14$-way classification task, where each input image may have multiple pathologies represented. We conduct $3$ main experiments (GAN-based augmentation, standard augmentation, no augmentation), and for each experiment we train using randomly sampled $\{1\%, 10\%, 50\%, 100\%\}$ subsets of the training dataset. We use a $90\%-5\%-5\%$ split for training/validation/testing. The same hold-out test set, consisting of 5\% of the original dataset, is used to evaluate all networks, while the random training/validation split takes place at train-time.

All networks are trained using binary cross entropy loss for up to $10$ epochs. While training for more epochs may lead to higher performance, computational constraints limit us to $10$ epochs per experiment. For each experiment, we conduct the following hyperparameter search:
\begin{itemize}
    \itemsep-.2em 
    \item Learning rate: $\{1e-2, 5e-3, 1e-3\}$
    \item Batch size: $\{16, 32, 64\}$
    \item Optimizer: \{Momentum, Adam\}
\end{itemize}
In the following sections, all AUCs reported are for the networks and hyperparameters that achieve the highest validation AUC.

\begin{figure*}
\centering
\includegraphics[width=1\linewidth]{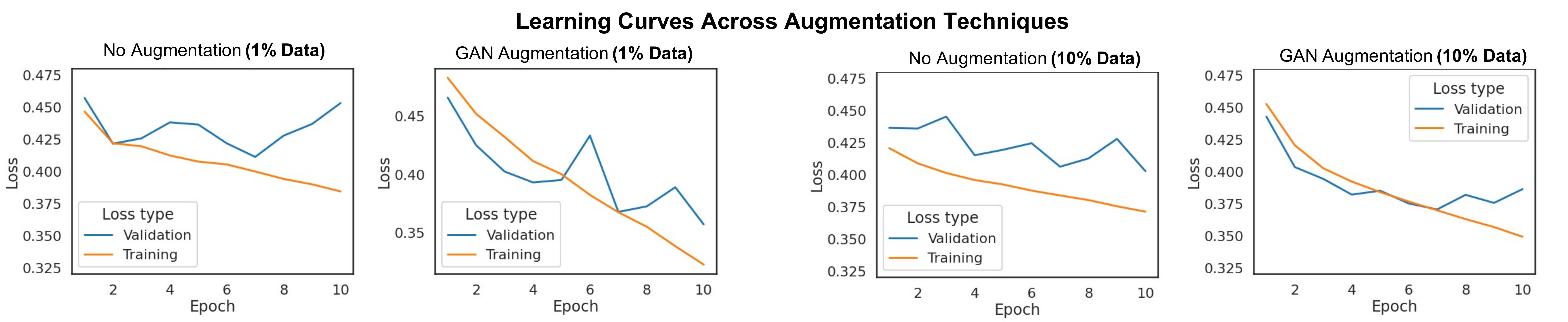}
\caption{Comparison of training and validation loss curves for no augmentation and GAN-based augmentation training}
\label{fig:losses}
\end{figure*}

\begin{figure}
\centering
\includegraphics[width=1\linewidth]{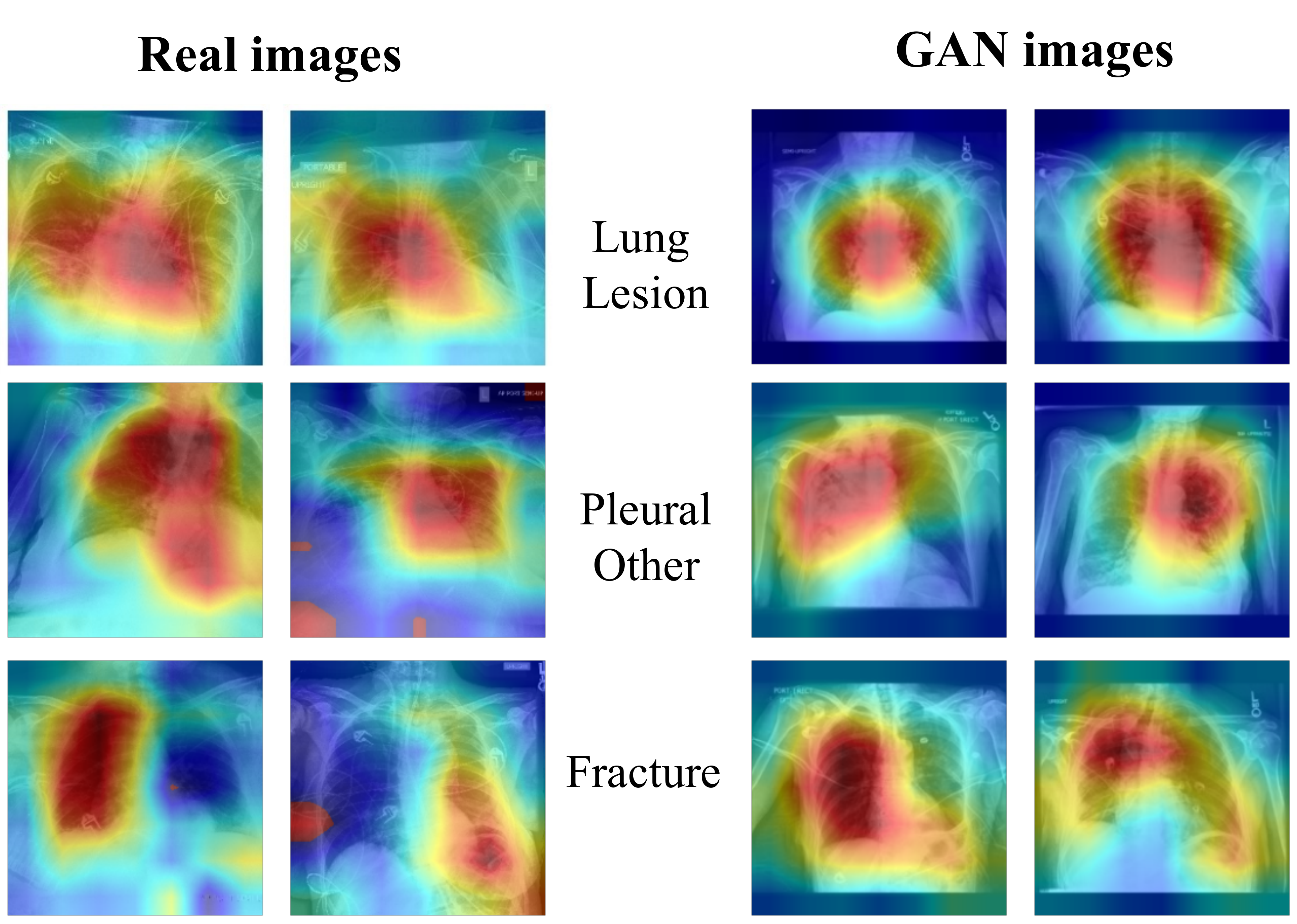}
\caption{Class Activation Maps (CAMs) produced by a network trained with GAN augmentation, for real images and GAN-generated images.}
\label{fig:CAMS}
\end{figure}

\subsection{Model performance}
Model performance via ROC-AUC score across the augmentation techniques and dataset regimens is detailed in Table \ref{AUC_table}. For the smallest data regimen (1\%) we see significant gains in performance when training with GAN-augmentation data as opposed to standard or no augmentation. This is evident across all pathologies, with a $0.03$ AUC gain between GAN augmentation and no augmentation for lung lesions and pleural other, and a $0.07$ AUC gain for fracture. Note that using standard augmentation also improves performance compared to no augmentation, however less so than GAN augmentation. Our results for the second-smallest data regimen (10\%) also demonstrate an improvement with GAN augementation across the pathologies. These positive results demonstrate that models trained with GAN augmentation outperform those trained with standard or no augmentation in small data regimens.

In the larger data regimens, we see either minimal or negative performance improvement for GAN and standard augmentation techniques as compared to no augmentation. For the data regimen of $100\%$, both augmentation techniques have a lesser performance than the original dataset for the lung lesion and fracture pathologies. However, in the pleural other pathology, we see a $0.006$ AUC gain from no augmentation to GAN augmentation. Since pleural other is the most imbalanced class of the three, this indicates that even at high data regimens, GAN augmentation can be useful for significant data imbalances.

Additionally, we observe that standard data augmentation has a worse performance than no augmentation across all pathologies in large data regimens. Our standard augmentation method uses flips; thus this result could be attributed to the presence of structures in the radiographs that lack left-right or up-down invariance. Alternate techniques could include blurring or saturation/hue jittering. The performance decrease for standard augmentation could also be alleviated with training standard-augmentation models for more epochs (we were unable to do so due to computational constraints). 
\vspace{-4.5mm}
\subsection{Comparison of learning curves}
In Figure \ref{fig:losses}, we compare learning curves when training with no augmentation and GAN augmentation. We find that GAN based augmentation alleviates overfitting in the low-data regimen of $1\%$. In the final epoch, we see a $0.03$ difference between training and validation loss in the GAN augmented model and a $0.06$ difference in the non-augmented model. In the $10\%$ regimen, the overfitting between the two models is comparable, with the GAN-augmented and non-augmented loss differences at $0.03$ and $0.04$, respectively. This is consistent with our AUC results, indicating that adding synthetic images can alleviate overfitting for very low data regimens, but does not necessarily help as the dataset size increases.

\subsection{Visualization of class activation maps}

We gain further qualitative insight regarding the use of GAN-based augmentation through class activation maps (CAMs) \cite{CAMPaper}. For a particular category, CAMs indicate the most discriminative image regions used to identify the category by projecting the weights of the output layer onto the final convolutional feature maps.

In Figure \ref{fig:CAMS}, we visualize CAMs for the DenseNet trained with GAN-augmentation on $100\%$ of the CheXpert training data, as this model had the highest overall performance. For each of the $3$ underrepresented pathologies, we visualize CAMs for real X-rays and GAN-generated X-rays. Note that for each pathology, the resultant heatmaps appear similar between real and GAN-generated images, indicating that the network activations may be similar between the two. The similarity of CAMs provides some evidence that the network has not overfit to real or GAN images, despite being trained on mostly GAN images for the imbalanced classes. 

Furthermore, for all CAMs, the highest activations appear localized to a particular region in each X-ray. The emergence of high activation at specific locations means that CAMs may provide useful, interpretable insight in a clinical setting regarding the parts of an X-ray that indicate specific pathologies.

\section{Conclusion}
In this work, we perform a comparison of GAN data-augmented models to standard augmented and non-augmented models across different data regimens. Our findings indicate that GAN-based data augmentation may be an effective tool for correcting class imbalanced medical datasets. We demonstrate AUC performance gains as compared to standard and no augmentation techniques across dataset sizes and pathologies. Additionally, we find that GAN-data augmented models overfit less in low-data regimens. From our results, we conclude that augmentation techniques are particularly effective across low-data regimens and show promise for correcting class imbalance in medical applications, where datasets are often both small and skewed. Note that this study assumes access to a GAN that has been pretrained on a large data set, which may not always be the case in a clinical setting. Future work may investigate whether the advantages of GAN-based augmentation also exist when using GANs trained on smaller datasets.

\bibliographystyle{ACM-Reference-Format}
\bibliography{sample-sigconf}


\begin{thebibliography}{23}


\ifx \showCODEN    \undefined \def \showCODEN     #1{\unskip}     \fi
\ifx \showDOI      \undefined \def \showDOI       #1{#1}\fi
\ifx \showISBNx    \undefined \def \showISBNx     #1{\unskip}     \fi
\ifx \showISBNxiii \undefined \def \showISBNxiii  #1{\unskip}     \fi
\ifx \showISSN     \undefined \def \showISSN      #1{\unskip}     \fi
\ifx \showLCCN     \undefined \def \showLCCN      #1{\unskip}     \fi
\ifx \shownote     \undefined \def \shownote      #1{#1}          \fi
\ifx \showarticletitle \undefined \def \showarticletitle #1{#1}   \fi
\ifx \showURL      \undefined \def \showURL       {\relax}        \fi
\providecommand\bibfield[2]{#2}
\providecommand\bibinfo[2]{#2}
\providecommand\natexlab[1]{#1}
\providecommand\showeprint[2][]{arXiv:#2}

\bibitem[\protect\citeauthoryear{Aggarwal~R}{Aggarwal~R}{2021}]%
        {imagingSurvey2}
\bibfield{author}{\bibinfo{person}{Martin G Ting DSW Karthikesalingam A King D
  Ashrafian H Darzi~A. Aggarwal~R, Sounderajah~V}.}
  \bibinfo{year}{2021}\natexlab{}.
\newblock \showarticletitle{Diagnostic accuracy of deep learning in medical
  imaging: a systematic review and meta-analysis}.
\newblock \bibinfo{journal}{\emph{NPJ Digit Med}}  \bibinfo{volume}{4}
  (\bibinfo{year}{2021}).
\newblock


\bibitem[\protect\citeauthoryear{Antoniou, Storkey, and Edwards}{Antoniou
  et~al\mbox{.}}{2018}]%
        {antoniou2018data}
\bibfield{author}{\bibinfo{person}{Anthreas Antoniou}, \bibinfo{person}{Amos
  Storkey}, {and} \bibinfo{person}{Harrison Edwards}.}
  \bibinfo{year}{2018}\natexlab{}.
\newblock \bibinfo{title}{Data Augmentation Generative Adversarial Networks}.
\newblock
\newblock
\urldef\tempurl%
\url{https://openreview.net/forum?id=S1Auv-WRZ}
\showURL{%
\tempurl}


\bibitem[\protect\citeauthoryear{Cohen, Hashir, Brooks, and Bertrand}{Cohen
  et~al\mbox{.}}{2020}]%
        {cohen2020limits}
\bibfield{author}{\bibinfo{person}{Joseph~Paul Cohen},
  \bibinfo{person}{Mohammad Hashir}, \bibinfo{person}{Rupert Brooks}, {and}
  \bibinfo{person}{Hadrien Bertrand}.} \bibinfo{year}{2020}\natexlab{}.
\newblock \showarticletitle{On the limits of cross-domain generalization in
  automated X-ray prediction}. In \bibinfo{booktitle}{\emph{Medical Imaging
  with Deep Learning}}.
\newblock
\urldef\tempurl%
\url{https://arxiv.org/abs/2002.02497}
\showURL{%
\tempurl}


\bibitem[\protect\citeauthoryear{Erhan, Bengio, Courville, and Vincent}{Erhan
  et~al\mbox{.}}{2009}]%
        {CAMPaper}
\bibfield{author}{\bibinfo{person}{D. Erhan}, \bibinfo{person}{Yoshua Bengio},
  \bibinfo{person}{Aaron~C. Courville}, {and} \bibinfo{person}{Pascal
  Vincent}.} \bibinfo{year}{2009}\natexlab{}.
\newblock \showarticletitle{Visualizing Higher-Layer Features of a Deep
  Network}.
\newblock


\bibitem[\protect\citeauthoryear{Goodfellow, Pouget-Abadie, Mirza, Xu,
  Warde-Farley, Ozair, Courville, and Bengio}{Goodfellow et~al\mbox{.}}{2014}]%
        {GoodfellowGAN}
\bibfield{author}{\bibinfo{person}{I. Goodfellow}, \bibinfo{person}{Jean
  Pouget-Abadie}, \bibinfo{person}{Mehdi Mirza}, \bibinfo{person}{B. Xu},
  \bibinfo{person}{David Warde-Farley}, \bibinfo{person}{S. Ozair},
  \bibinfo{person}{Aaron~C. Courville}, {and} \bibinfo{person}{Yoshua Bengio}.}
  \bibinfo{year}{2014}\natexlab{}.
\newblock \showarticletitle{Generative Adversarial Networks}.
\newblock \bibinfo{journal}{\emph{ArXiv}}  \bibinfo{volume}{abs/1406.2661}
  (\bibinfo{year}{2014}).
\newblock


\bibitem[\protect\citeauthoryear{Han, Nebelung, Haarburger, Horst, Reinartz,
  Merhof, Kiessling, Schulz, and Truhn}{Han et~al\mbox{.}}{2019}]%
        {conditionalGAN}
\bibfield{author}{\bibinfo{person}{Tianyu Han}, \bibinfo{person}{Sven
  Nebelung}, \bibinfo{person}{Christoph Haarburger}, \bibinfo{person}{Nicolas
  Horst}, \bibinfo{person}{Sebastian Reinartz}, \bibinfo{person}{Dorit Merhof},
  \bibinfo{person}{Fabian Kiessling}, \bibinfo{person}{Volkmar Schulz}, {and}
  \bibinfo{person}{Daniel Truhn}.} \bibinfo{year}{2019}\natexlab{}.
\newblock \showarticletitle{Breaking Medical Data Sharing Boundaries by
  Employing Artificial Radiographs}.
\newblock \bibinfo{journal}{\emph{bioRxiv}} (\bibinfo{year}{2019}).
\newblock
\urldef\tempurl%
\url{https://doi.org/10.1101/841619}
\showDOI{\tempurl}
\showeprint{https://www.biorxiv.org/content/early/2019/11/14/841619.full.pdf}


\bibitem[\protect\citeauthoryear{Irvin, Rajpurkar, Ko, Yu, Ciurea-Ilcus, Chute,
  Marklund, Haghgoo, Ball, Shpanskaya, Seekins, Mong, Halabi, Sandberg, Jones,
  Larson, Langlotz, Patel, Lungren, and Ng}{Irvin et~al\mbox{.}}{2019}]%
        {Irvin2019CheXpertAL}
\bibfield{author}{\bibinfo{person}{Jeremy~A. Irvin}, \bibinfo{person}{Pranav
  Rajpurkar}, \bibinfo{person}{M. Ko}, \bibinfo{person}{Yifan Yu},
  \bibinfo{person}{Silviana Ciurea-Ilcus}, \bibinfo{person}{Chris Chute},
  \bibinfo{person}{H. Marklund}, \bibinfo{person}{Behzad Haghgoo},
  \bibinfo{person}{Robyn~L. Ball}, \bibinfo{person}{K. Shpanskaya},
  \bibinfo{person}{J. Seekins}, \bibinfo{person}{D. Mong}, \bibinfo{person}{S.
  Halabi}, \bibinfo{person}{J. Sandberg}, \bibinfo{person}{R. Jones},
  \bibinfo{person}{D. Larson}, \bibinfo{person}{C. Langlotz},
  \bibinfo{person}{B. Patel}, \bibinfo{person}{M. Lungren}, {and}
  \bibinfo{person}{A. Ng}.} \bibinfo{year}{2019}\natexlab{}.
\newblock \showarticletitle{CheXpert: A Large Chest Radiograph Dataset with
  Uncertainty Labels and Expert Comparison}. In
  \bibinfo{booktitle}{\emph{AAAI}}.
\newblock


\bibitem[\protect\citeauthoryear{Japkowicz and Stephen}{Japkowicz and
  Stephen}{2002}]%
        {japkowicz2002class}
\bibfield{author}{\bibinfo{person}{Nathalie Japkowicz} {and}
  \bibinfo{person}{Shaju Stephen}.} \bibinfo{year}{2002}\natexlab{}.
\newblock \showarticletitle{The class imbalance problem: A systematic study}.
\newblock \bibinfo{journal}{\emph{Intelligent data analysis}}
  \bibinfo{volume}{6}, \bibinfo{number}{5} (\bibinfo{year}{2002}),
  \bibinfo{pages}{429--449}.
\newblock


\bibitem[\protect\citeauthoryear{Krawczyk}{Krawczyk}{2016}]%
        {krawczyk2016learning}
\bibfield{author}{\bibinfo{person}{Bartosz Krawczyk}.}
  \bibinfo{year}{2016}\natexlab{}.
\newblock \showarticletitle{Learning from imbalanced data: open challenges and
  future directions}.
\newblock \bibinfo{journal}{\emph{Progress in Artificial Intelligence}}
  \bibinfo{volume}{5}, \bibinfo{number}{4} (\bibinfo{year}{2016}),
  \bibinfo{pages}{221--232}.
\newblock


\bibitem[\protect\citeauthoryear{Liu, Song, Liu, and Zhang}{Liu
  et~al\mbox{.}}{2021}]%
        {segmentationSurvey}
\bibfield{author}{\bibinfo{person}{Xiangbin Liu}, \bibinfo{person}{Liping
  Song}, \bibinfo{person}{Shuai Liu}, {and} \bibinfo{person}{Yudong Zhang}.}
  \bibinfo{year}{2021}\natexlab{}.
\newblock \showarticletitle{A Review of Deep-Learning-Based Medical Image
  Segmentation Methods}.
\newblock \bibinfo{journal}{\emph{Sustainability}} \bibinfo{volume}{13},
  \bibinfo{number}{3} (\bibinfo{year}{2021}).
\newblock
\showISSN{2071-1050}
\urldef\tempurl%
\url{https://doi.org/10.3390/su13031224}
\showDOI{\tempurl}


\bibitem[\protect\citeauthoryear{Mariani, Scheidegger, Istrate, Bekas, and
  Malossi}{Mariani et~al\mbox{.}}{2018}]%
        {Mariani2018BAGANDA}
\bibfield{author}{\bibinfo{person}{G. Mariani}, \bibinfo{person}{F.
  Scheidegger}, \bibinfo{person}{R. Istrate}, \bibinfo{person}{C. Bekas}, {and}
  \bibinfo{person}{A. Malossi}.} \bibinfo{year}{2018}\natexlab{}.
\newblock \showarticletitle{BAGAN: Data Augmentation with Balancing GAN}.
\newblock \bibinfo{journal}{\emph{ArXiv}}  \bibinfo{volume}{abs/1803.09655}
  (\bibinfo{year}{2018}).
\newblock


\bibitem[\protect\citeauthoryear{Pham, Le, Ngo, Tran, and Nguyen}{Pham
  et~al\mbox{.}}{2021}]%
        {Pham2021InterpretingCX}
\bibfield{author}{\bibinfo{person}{Hieu~H. Pham}, \bibinfo{person}{Tung~T. Le},
  \bibinfo{person}{D. Ngo}, \bibinfo{person}{Dat~Q. Tran}, {and}
  \bibinfo{person}{H.~Q. Nguyen}.} \bibinfo{year}{2021}\natexlab{}.
\newblock \showarticletitle{Interpreting Chest X-rays via CNNs that Exploit
  Hierarchical Disease Dependencies and Uncertainty Labels}.
\newblock \bibinfo{journal}{\emph{Neurocomputing}}  \bibinfo{volume}{437}
  (\bibinfo{year}{2021}), \bibinfo{pages}{186--194}.
\newblock


\bibitem[\protect\citeauthoryear{Pham, Le, Tran, Ngo, and Nguyen}{Pham
  et~al\mbox{.}}{2019}]%
        {Densenet_for_XRay}
\bibfield{author}{\bibinfo{person}{Hieu~H. Pham}, \bibinfo{person}{Tung~T. Le},
  \bibinfo{person}{Dat~Q. Tran}, \bibinfo{person}{D. Ngo}, {and}
  \bibinfo{person}{H. Nguyen}.} \bibinfo{year}{2019}\natexlab{}.
\newblock \showarticletitle{Interpreting chest X-rays via CNNs that exploit
  disease dependencies and uncertainty labels}.
\newblock \bibinfo{journal}{\emph{ArXiv}}  \bibinfo{volume}{abs/1911.06475}
  (\bibinfo{year}{2019}).
\newblock


\bibitem[\protect\citeauthoryear{Qasim, Ezhov, Shit, Schoppe, Paetzold,
  Sekuboyina, Kofler, Lipkov{\'a}, Li, and Menze}{Qasim et~al\mbox{.}}{2020}]%
        {Qasim2020RedGANAC}
\bibfield{author}{\bibinfo{person}{A.~B. Qasim}, \bibinfo{person}{Ivan Ezhov},
  \bibinfo{person}{S. Shit}, \bibinfo{person}{Oliver Schoppe},
  \bibinfo{person}{Johannes~C. Paetzold}, \bibinfo{person}{A. Sekuboyina},
  \bibinfo{person}{Florian Kofler}, \bibinfo{person}{Jana Lipkov{\'a}},
  \bibinfo{person}{Hongwei Li}, {and} \bibinfo{person}{B. Menze}.}
  \bibinfo{year}{2020}\natexlab{}.
\newblock \showarticletitle{Red-GAN: Attacking class imbalance via conditioned
  generation. Yet another medical imaging perspective}.
\newblock \bibinfo{journal}{\emph{ArXiv}}  \bibinfo{volume}{abs/2004.10734}
  (\bibinfo{year}{2020}).
\newblock


\bibitem[\protect\citeauthoryear{Rajpurkar, Irvin, Ball, Zhu, Yang, Mehta,
  Duan, Ding, Bagul, Langlotz, Patel, Yeom, Shpanskaya, Blankenberg, Seekins,
  Amrhein, Mong, Halabi, Zucker, Ng, and Lungren}{Rajpurkar
  et~al\mbox{.}}{2018}]%
        {Rajpurkar2018DeepLF}
\bibfield{author}{\bibinfo{person}{Pranav Rajpurkar},
  \bibinfo{person}{Jeremy~A. Irvin}, \bibinfo{person}{Robyn~L. Ball},
  \bibinfo{person}{Kaylie Zhu}, \bibinfo{person}{B. Yang},
  \bibinfo{person}{Hershel Mehta}, \bibinfo{person}{T. Duan},
  \bibinfo{person}{D. Ding}, \bibinfo{person}{Aarti Bagul}, \bibinfo{person}{C.
  Langlotz}, \bibinfo{person}{B. Patel}, \bibinfo{person}{K. Yeom},
  \bibinfo{person}{K. Shpanskaya}, \bibinfo{person}{F. Blankenberg},
  \bibinfo{person}{J. Seekins}, \bibinfo{person}{T. Amrhein},
  \bibinfo{person}{D. Mong}, \bibinfo{person}{S. Halabi}, \bibinfo{person}{Evan
  Zucker}, \bibinfo{person}{A. Ng}, {and} \bibinfo{person}{M. Lungren}.}
  \bibinfo{year}{2018}\natexlab{}.
\newblock \showarticletitle{Deep learning for chest radiograph diagnosis: A
  retrospective comparison of the CheXNeXt algorithm to practicing
  radiologists}.
\newblock \bibinfo{journal}{\emph{PLoS Medicine}}  \bibinfo{volume}{15}
  (\bibinfo{year}{2018}).
\newblock


\bibitem[\protect\citeauthoryear{Rashid, Tanveer, and Khan}{Rashid
  et~al\mbox{.}}{2019}]%
        {Rashid_skin_lesion_GAN_augmentation}
\bibfield{author}{\bibinfo{person}{Haroon Rashid}, \bibinfo{person}{M.
  Tanveer}, {and} \bibinfo{person}{H. Khan}.} \bibinfo{year}{2019}\natexlab{}.
\newblock \showarticletitle{Skin Lesion Classification Using GAN based Data
  Augmentation}.
\newblock \bibinfo{journal}{\emph{2019 41st Annual International Conference of
  the IEEE Engineering in Medicine and Biology Society (EMBC)}}
  (\bibinfo{year}{2019}), \bibinfo{pages}{916--919}.
\newblock


\bibitem[\protect\citeauthoryear{Rice, Wong, and Kolter}{Rice
  et~al\mbox{.}}{2020}]%
        {overfitting}
\bibfield{author}{\bibinfo{person}{Leslie Rice}, \bibinfo{person}{Eric Wong},
  {and} \bibinfo{person}{Zico Kolter}.} \bibinfo{year}{2020}\natexlab{}.
\newblock \showarticletitle{Overfitting in adversarially robust deep learning}.
  In \bibinfo{booktitle}{\emph{Proceedings of the 37th International Conference
  on Machine Learning}} \emph{(\bibinfo{series}{Proceedings of Machine Learning
  Research}, Vol.~\bibinfo{volume}{119})},
  \bibfield{editor}{\bibinfo{person}{Hal~Daumé III} {and}
  \bibinfo{person}{Aarti Singh}} (Eds.). \bibinfo{publisher}{PMLR},
  \bibinfo{pages}{8093--8104}.
\newblock
\urldef\tempurl%
\url{http://proceedings.mlr.press/v119/rice20a.html}
\showURL{%
\tempurl}


\bibitem[\protect\citeauthoryear{Sampath, Maurtua, Mart{\'i}n, and
  Gutierrez}{Sampath et~al\mbox{.}}{2020}]%
        {Sampath2020ASO}
\bibfield{author}{\bibinfo{person}{Vignesh Sampath}, \bibinfo{person}{I.
  Maurtua}, \bibinfo{person}{Juan Jos{\'e}~Aguilar Mart{\'i}n}, {and}
  \bibinfo{person}{Aitor Gutierrez}.} \bibinfo{year}{2020}\natexlab{}.
\newblock \showarticletitle{A Survey on Generative Adversarial Networks for
  imbalance problems in computer vision tasks}.
\newblock


\bibitem[\protect\citeauthoryear{Sandfort, Yan, Pickhardt, and
  Summers}{Sandfort et~al\mbox{.}}{2019}]%
        {CycleGAN_CT_scan}
\bibfield{author}{\bibinfo{person}{V. Sandfort}, \bibinfo{person}{Ke Yan},
  \bibinfo{person}{P. Pickhardt}, {and} \bibinfo{person}{R. Summers}.}
  \bibinfo{year}{2019}\natexlab{}.
\newblock \showarticletitle{Data augmentation using generative adversarial
  networks (CycleGAN) to improve generalizability in CT segmentation tasks}.
\newblock \bibinfo{journal}{\emph{Scientific Reports}}  \bibinfo{volume}{9}
  (\bibinfo{year}{2019}).
\newblock


\bibitem[\protect\citeauthoryear{Shorten and Khoshgoftaar}{Shorten and
  Khoshgoftaar}{2019}]%
        {Shorten2019ASO}
\bibfield{author}{\bibinfo{person}{Connor Shorten} {and} \bibinfo{person}{T.
  Khoshgoftaar}.} \bibinfo{year}{2019}\natexlab{}.
\newblock \showarticletitle{A survey on Image Data Augmentation for Deep
  Learning}.
\newblock \bibinfo{journal}{\emph{Journal of Big Data}}  \bibinfo{volume}{6}
  (\bibinfo{year}{2019}), \bibinfo{pages}{1--48}.
\newblock


\bibitem[\protect\citeauthoryear{Stephen, Sain, Maduh, and Jeong}{Stephen
  et~al\mbox{.}}{2019}]%
        {Stephen2019AnED}
\bibfield{author}{\bibinfo{person}{O. Stephen}, \bibinfo{person}{M. Sain},
  \bibinfo{person}{U.~J. Maduh}, {and} \bibinfo{person}{Do-Un Jeong}.}
  \bibinfo{year}{2019}\natexlab{}.
\newblock \showarticletitle{An Efficient Deep Learning Approach to Pneumonia
  Classification in Healthcare}.
\newblock \bibinfo{journal}{\emph{Journal of Healthcare Engineering}}
  \bibinfo{volume}{2019} (\bibinfo{year}{2019}).
\newblock


\bibitem[\protect\citeauthoryear{Stutz, Hein, and Schiele}{Stutz
  et~al\mbox{.}}{2019}]%
        {stutz2019disentangling}
\bibfield{author}{\bibinfo{person}{David Stutz}, \bibinfo{person}{Matthias
  Hein}, {and} \bibinfo{person}{Bernt Schiele}.}
  \bibinfo{year}{2019}\natexlab{}.
\newblock \showarticletitle{Disentangling adversarial robustness and
  generalization}. In \bibinfo{booktitle}{\emph{Proceedings of the IEEE/CVF
  Conference on Computer Vision and Pattern Recognition}}.
  \bibinfo{pages}{6976--6987}.
\newblock


\bibitem[\protect\citeauthoryear{Zhou, Greenspan, Davatzikos, Duncan,
  Van~Ginneken, Madabhushi, Prince, Rueckert, and Summers}{Zhou
  et~al\mbox{.}}{2021}]%
        {imagingSurvey}
\bibfield{author}{\bibinfo{person}{S.~Kevin Zhou}, \bibinfo{person}{Hayit
  Greenspan}, \bibinfo{person}{Christos Davatzikos}, \bibinfo{person}{James~S.
  Duncan}, \bibinfo{person}{Bram Van~Ginneken}, \bibinfo{person}{Anant
  Madabhushi}, \bibinfo{person}{Jerry~L. Prince}, \bibinfo{person}{Daniel
  Rueckert}, {and} \bibinfo{person}{Ronald~M. Summers}.}
  \bibinfo{year}{2021}\natexlab{}.
\newblock \showarticletitle{A Review of Deep Learning in Medical Imaging:
  Imaging Traits, Technology Trends, Case Studies With Progress Highlights, and
  Future Promises}.
\newblock \bibinfo{journal}{\emph{Proc. IEEE}} \bibinfo{volume}{109},
  \bibinfo{number}{5} (\bibinfo{year}{2021}), \bibinfo{pages}{820--838}.
\newblock
\urldef\tempurl%
\url{https://doi.org/10.1109/JPROC.2021.3054390}
\showDOI{\tempurl}


\end{thebibliography}










\end{document}